\documentclass[ notitlepage, reprint, amsmath,amssymb, aps, prl ]{revtex4-2}

\usepackage{xcolor}
\usepackage{graphicx}% Include figure files
\usepackage{dcolumn}% Align table columns on decimal point
\usepackage{bm}% bold math
\usepackage{multirow}
\usepackage[T1]{fontenc}
\usepackage{times}

\begin{document}

\title {
  Dichotomy of Electron-Phonon Coupling in Graphene Moir\'e Flat Bands
}

\author{Young Woo Choi}
\author{Hyoung Joon Choi}
\email{h.j.choi@yonsei.ac.kr}
\affiliation{Department of Physics, Yonsei University, Seoul 03722, Korea}

\begin{abstract}
  Graphene moir\'e superlattices are outstanding platforms to study correlated
  electron physics and superconductivity with exceptional tunability.
  However, robust superconductivity has been measured only in magic-angle
  twisted bilayer graphene (MA-TBG) and magic-angle twisted trilayer graphene (MA-TTG).
  The absence of a superconducting phase in certain moir\'e flat bands
  raises a question on the superconducting mechanism.
  In this work, we investigate electronic structure and electron-phonon
  coupling in graphene moir\'e superlattices based on atomistic
  calculations.
  We show that electron-phonon coupling strength $\lambda$ is dramatically
  different among graphene moir\'e flat bands.
  The total strength $\lambda$ is very large ($\lambda>1$) for MA-TBG and
  MA-TTG, both of which display robust superconductivity in experiments.
  However, $\lambda$ is an order of magnitude smaller in twisted double bilayer
  graphene (TDBG) and twisted monolayer-bilayer graphene (TMBG) where
  superconductivity is reportedly rather weak or absent.
  We find that the Bernal-stacked layers in TDBG and TMBG induce
  sublattice polarization in the flat-band states,
  suppressing intersublattice electron-phonon matrix elements.
  We also obtain the nonadiabatic superconducting transition temperature $T_c$ that matches well with
  the experimental results.
  Our results clearly show a correlation between strong electron-phonon
  coupling and experimental observations of robust superconductivity.
\end{abstract}

\maketitle

Moir\'e materials have emerged as
precisely tunable platforms to explore fascinating physical phenomena
\cite{andrei_2021}.
For example, twisted bilayer graphene (TBG) was predicted
to have nearly flat bands at certain "magic angles" (MAs)
\cite{bistritzer_2011,lopesdossantos_2007,shallcross_2010,suarezmorell_2010},
and experimentally shown to host correlated insulators and superconductivity
\cite{cao_2018,cao_2018a}.
Subsequently, various interaction-driven phases have been realized not only in
magic-angle twisted bilayer graphene (MA-TBG)
\cite{choi_2021b,nuckolls_2020,polshyn_2020,serlin_2020,wong_2020,zondiner_2020}
but also in other graphene moir\'e superlattices
\cite{chen_2019,liu_2020,cao_2020b,chen_2020b,hao_2021a,park_2021,chen_2021}.
In addition to electronic properties, extensive aspects of moir\'e physics have
been actively explored, such as moir\'e excitons \cite{alexeev_2019,tran_2019}
and atomic-structure and phonon reconstructions
\cite{yoo_2019,quan_2021,gadelha_2021}.

While correlated insulating states are observed in many graphene moir\'e
superlattices having flat bands,
MA-TBG has been the only system to show robust superconductivity
until recent experiments added magic-angle alternating-twist trilayer graphene
(TTG) to the list \cite{cao_2018a,hao_2021a,park_2021}.
In other systems, such as twisted double bilayer graphene (TDBG)
and twisted monolayer-bilayer graphene (TMBG),
superconducting phase appears to be rather weak or absent
\cite{liu_2020,cao_2020b,chen_2020a}.

Naturally, the absence of robust superconductivity in some graphene moir\'e
flat bands raises more questions on the superconducting mechanism.
In particular, theoretical studies on MA-TBG have suggested that MA-TBG has
strong electron-phonon coupling and phonon-mediated superconductivity is a
strong candidate \cite{wu_2018a,choi_2018,peltonen_2018,lian_2019}. Also,
strong electron-phonon coupling in MA-TBG is evidenced in experiments
\cite{polshyn_2019,gadelha_2021}.
However, whether all the flat-band states in graphene moir\'e superlattices
have such strong electron-phonon coupling is still unanswered, which
has an important implication for the superconducting mechanism.

\begin{figure}
  \includegraphics[width=8.6cm]{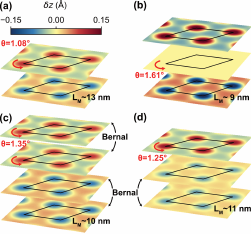}
  \caption {
    Atomic relaxation patterns of
    (a) TBG with $\theta=1.08^\circ$,
    (b) TTG with an alternating twist angle $\theta=1.61^\circ$,
    (c) TDBG with $\theta=1.35^\circ$, and
    (d) TMBG with $\theta=1.25^\circ$.
    $\delta z$ is the deviation of atomic positions in the out-of-plane
    direction from the average value within each layer. Topmost two layers in
    (c) and bottom two layers in (c) and (d) are Bernal-stacked without twist.
  }
  \label{fig:fig1}
\end{figure}

\begin{figure*}
  \includegraphics[width=\textwidth]{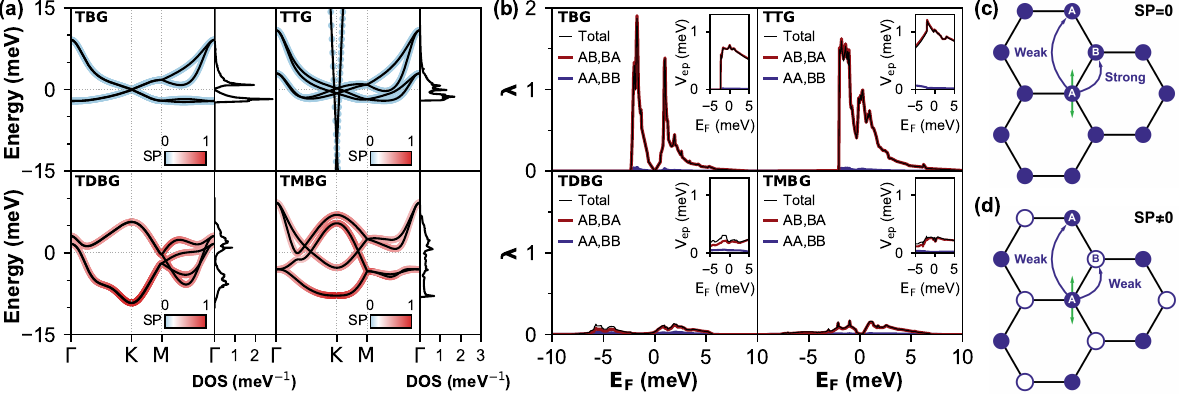}
  \caption {
    (a) Band structure and density of states (DOS) per spin and (b)
    electron-phonon coupling strength $\lambda$ as a function of the Fermi
    energies $E_F$ in (upper left) TBG with $\theta=1.08^\circ$,
    (upper right) TTG with $\theta=1.61^\circ$,
    (lower left) TDBG with $\theta=1.35^\circ$,
    and (lower right) TMBG with $\theta=1.25^\circ$.
    In (a), colored circles represent sublattice polarization (SP).
    TBG and TTG have exactly zero sublattice polarization.
    In (b), black lines show the total coupling strength.
    Red (blue) lines represent contributions of the inter- (intra-) sublattice
    electron-phonon coupling.
    In each case, the inset shows the average electron-phonon matrix element
    $V_{\mathrm{ep}}=\lambda/N_F$, where $N_F$ is the DOS per spin at $E_F$.
    (c),(d) Illustrations of how sublattice polarization suppresses
    intersublattice electron-phonon matrix elements. Blue arrows indicate
    electron-phonon matrix elements induced by phonon displacements denoted by
    green arrows.
    When sublattice polarization is nonzero in (d), the nearest-neighbor
    electron-phonon matrix elements are weakened.
  }
  \label{fig:fig2}
\end{figure*}

In this work, we investigate electronic structure and electron-phonon coupling
of graphene moir\'e superlattices based on atomistic calculations.
We show that electron-phonon coupling is strong ($\lambda>1$) for TBG and TTG
at their magic angles, but it is an order of magnitude weaker for TDBG and
TMBG.
We analyze such difference in $\lambda$ in terms of both density of states and
matrix-element effects. We find that Bernal-stacked layers in TDBG and TMBG
induce sublattice polarization in flat-band states, suppressing
intersublattice electron-phonon matrix elements.
Regardless of the total coupling strength, characteristic phonon modes are the same
for all systems. We also calculate effects of vertical electric fields on
electronic structure and electron-phonon coupling.
Our calculations show that a correlation exists between strong electron-phonon
coupling and experimental observations of robust superconductivity.

We consider four graphene moir\'e superlattices at their respective
magic angle, which is defined by the angle of the minimum bandwidth:
TBG with $\theta=1.08^\circ$,
TTG with an alternating twist angle $\theta=1.61^\circ$,
TDBG with $\theta=1.35^\circ$, and
TMBG with $\theta=1.25^\circ$.
TBG consists of two graphene layers with a twist,
TTG is a three-layer system where only the middle layer is twisted,
TDBG has two sets of Bernal-stacked bilayer graphene with a twist between them,
and TMBG is built by twisting a monolayer on the top of Bernal-stacked bilayer
graphene.

We adopt atomistic approaches to calculate electrons and phonons in
graphene moir\'e superlattices \cite{choi_2018}.
First, we calculate structural relaxations induced by variation of the stacking
registry within moir\'e supercells, which have crucial effects on
electronic structure of moir\'e flat bands \cite{nam_2017,yoo_2019}.
Equilibrium positions of all the carbon atoms are obtained by minimizing
the sum of in-plane elastic energy and interlayer van der Waals binding energy
\cite{wirtz_2004,kolmogorov_2005}.
Then, electron states are obtained by diagonalizing atomistic tight-binding
Hamiltonians with the Slater-Koster-type hopping integral
parameterized for graphitic systems \cite{nakanishi_2001,moon_2012}.
We calculate all the phonon modes in moir\'e supercells by diagonalizing
dynamical matrices built from atomic force constants, which are the second
derivatives of our total-energy function.
With electron and phonon eigenstates, we compute electron-phonon matrix elements
from changes in hopping amplitudes due to atomic displacements of phonon modes.
From the above quantities, we can obtain electron-phonon coupling strength
$\lambda$ and the Eliashberg function $\alpha^2F(\omega)$
(See the Supplemental Material \cite{supplemental} for detailed descriptions of our methods).

Figure~\ref{fig:fig1} shows our results of atomic relaxation patterns.
$\delta z$ is the deviation of atomic positions in the out-of-plane direction
from the average value of each layer.
The average interlayer distances are about 3.40 \AA~between twisted layers and
3.35 \AA~between Bernal-stacked layers in TDBG and TMBG.
In all systems, $\delta z$ is largest at $AA$ stacking regions and has the
opposite sign between twisted layers, except for TTG where the middle layer
has zero $\delta z$ due to the symmetry and the other layers have large
$\delta z$ in compensation.
In our calculation, in-plane relaxations also occur in such a way to reduce the
area of $AA$-stacked regions.

\begin{figure}
  \includegraphics[width=8.6cm]{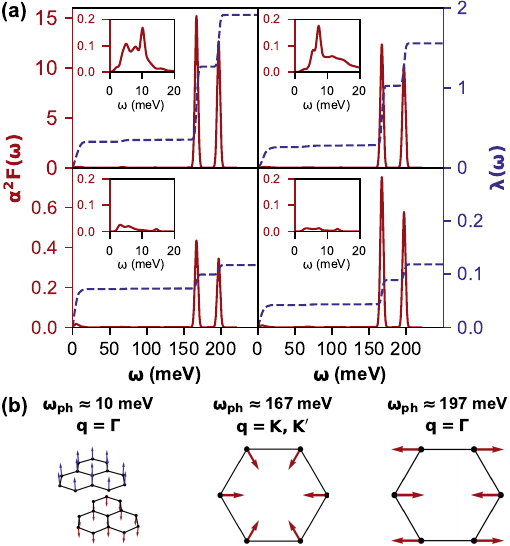}
  \caption {
    (a) Eliashberg function $\alpha^2F(\omega)$, shown in red, at the
    half-filling energy of the hole-side in (upper left) TBG and (upper right)
    TTG, and the electron-side in (lower left) TDBG and (lower right) TMBG.
    Dashed blue lines denote frequency-integrated electron-phonon coupling
    strength $\lambda(\omega)=2\int^{\omega}_{0} \alpha^2 F(\omega')/\omega'
    d\omega'$. Insets show low-frequency ranges of $\alpha^2 F(\omega)$.
    (b) Phonon frequencies and momenta of the characteristic modes.
  }
  \label{fig:fig3}
\end{figure}

Figure~\ref{fig:fig2}(a) shows our calculated band structures and density of
states (DOS) per spin.
All the four systems have nearly flat bands and large DOS at their Fermi levels.
Flat bands in TBG are the most archetypical in that
nearly flat Dirac cones are located at the corners of the moir\'e Brillouin zone
and isolated from the higher-energy bands.
In TTG, the highly dispersing Dirac cone coexists with moir\'e flat bands.
It comes from the outer graphene layers and is decoupled from the flat bands.
In TDBG and TMBG, Dirac points at $K$ points are gapped because the absence of
the inversion symmetry brings sublattice asymmetry.
Our electronic structures are consistent with previous theoretical studies
\cite{choi_2018,choi_2019,carr_2020b,angeli_2019,carr_2018}.

While TBG, TTG, TDBG, and TMBG at their respective magic angle have flat bands
and large DOS at the Fermi level, we find a clear distinction of
electron-phonon coupling strength $\lambda$ between the first two and last two
systems.
Figure~\ref{fig:fig2}(b) shows the total electron-phonon coupling strength
of each system as a function of the Fermi energy.
The most notable feature is that $\lambda$ is very strong in TBG and TTG with
the maximum value reaching over 1, but it is an order of magnitude weaker in TDBG
and TMBG. For TDBG and TMBG, $\lambda$ is less than 0.2 irrespective of the Fermi energy.

Such stark contrast in $\lambda$ originates partly from the
difference in the electronic density of states, but more crucially from the
suppression of electron-phonon matrix elements in TDBG and TMBG.
Insets in Fig.~\ref{fig:fig2}(b) show
the average of electron-phonon matrix elements $V_{\mathrm{ep}}=\lambda/N_F$,
where $N_F$ is the electronic density of states at the Fermi energy $E_F$.
While $V_{\mathrm{ep}}$ in TBG and TTG is about 0.75 and 1.0 meV, respectively,
it is below 0.3 meV in TDBG and TMBG.

To explain why $V_{\mathrm{ep}}$ is suppressed in TDBG and TMBG,
we introduce sublattice polarization (SP) which measures the imbalance of
sublattice weights of an electron state.
For a given electron state $\psi_{n\mathbf{k}}$, we define SP for each layer
$l$ as
\begin{equation}
\mathrm{SP}_{l}(\psi_{n\mathbf{k}}) =
\sum_{i\in A} |c_{n\mathbf{k},i}|^2 -
\sum_{j\in B} |c_{n\mathbf{k},j}|^2,
\end{equation}
where $c_{n\mathbf{k},i}$ is the tight-binding coefficient of $\psi_{n\mathbf{k}}$
for an orbital centered at an atomic site $i$ and $A,B$ indicate two different sublattices.
Then, the total SP, which is represented by colored circles in
Fig.~\ref{fig:fig2}(a), is $\mathrm{SP}\left(\psi_{n\mathbf{k}}\right)
=\sum_{l} \left|\mathrm{SP}_{l}(\psi_{n\mathbf{k}})\right|$.
In TBG and TTG,
the SP is exactly zero for all the electron states so they have
the equal weights on two sublattices.
In contrast,
TDBG and TMBG, which both have Bernal-stacked layers,
have nonzero SP and electrons have different sublattice weights
within each layer,
with signs of SP$_{l}$ alternating for different layers.

The presence of nonzero SP in the electronic structure of TDBG and TMBG
critically weakens electron-phonon coupling strength.
To illustrate this, we analyze the total electron-phonon coupling strength
in terms of sublattice-dependent contributions.
Figure~\ref{fig:fig2}(b) shows the total electron-phonon coupling strength $\lambda$, and
intersublattice ($\lambda^{AB}+\lambda^{BA}$) and intrasublattice ($\lambda^{AA}+\lambda^{BB}$) contributions
as a function of the Fermi energy
(see the Supplemental Material \cite{supplemental} for the formulas for
inter- and intrasublattice $\lambda$).
We find that, in all cases, the magnitude of the inter-sublattice contributions dominates
the total coupling strength.
This is because the strongest contribution comes from the electron-phonon
matrix elements between the nearest neighbors, which belong to different
sublattices [Fig.~\ref{fig:fig2}(c)].
In TBG and TTG, where SP is zero,
electron wave functions have the same weights on both sublattices
and the intersublattice matrix elements are strong.
On the other hand, in TDBG and TMBG, nonzero SP suppresses the nearest-neighbor matrix
elements [Fig.~\ref{fig:fig2}(c)], and the electron-phonon coupling becomes very weak.

\begin{table}[b]
\caption{\label{tab:table1}%
  Mode-resolved electron-phonon coupling strength $\lambda_i$,
  half bandwidth $D$,
  and nonadiabatic superconducting transition temperature $T_c$
  at the half-filling Fermi energy of the hole- (electron-) side flat bands
  for TBG and TTG (TDBG and TMBG).
  The half bandwidths are calculated as the difference between the Fermi energy
  and the band edge.
  $\mu$ is the dimensionless Coulomb potential.
}
\begin{ruledtabular}
\begin{tabular}{ccccc}
  \multirow[c]{2}{*}{$\omega_i$ (meV)} & \multicolumn{4}{c}{$\lambda_i$} \\
                                         \cline{2-5}
                                         & TBG & TTG & TDBG & TMBG \\
\hline
  10  & 0.297 & 0.233 & 0.064 & 0.037 \\
  167 & 0.914 & 0.743 & 0.026 & 0.045 \\
  197 & 0.648 & 0.532 & 0.018 & 0.030 \\
\hline
\hline
  \multicolumn{1}{c}{$D$ (meV)            } & 0.53 & 0.67 & 3.4 & 6.7 \\
  \multicolumn{1}{c}{$T_c(\mu=0.05) $ (K) } & 3.45 & 3.76 & $10^{-7}$ & $10^{-6}$ \\
  \multicolumn{1}{c}{$T_c(\mu=0.15) $ (K) } & 3.33 & 3.55 & 0.0 & 0.0
\end{tabular}

\end{ruledtabular}
\end{table}

Now, we analyze characteristic phonon modes that contribute to the total
coupling strength. Figure~\ref{fig:fig3}(a) shows the Eliashberg functions
$\alpha^2F(\omega)$ and frequency-integrated electron-phonon coupling strength
$\lambda(\omega)=2\int^{\omega}_{0} \alpha^2 F(\omega')/\omega' d\omega'$ of
TBG and TTG (TDBG and TMBG) at the half-filling energy of the hole- (electron-)
side. Regardless of the total coupling strength, characteristic phonon modes are the
same for all systems. The largest portion of the total coupling strength comes
from the in-plane optical modes near $\omega$ = 167 (197) meV with phonon
momentums at $\mathbf{q}=\mathbf{K},\mathbf{K'}$ ($\mathbf{\Gamma}$). In
addition, the interlayer breathing modes near $\omega$ = 10 meV, shown in the insets of
Fig.~\ref{fig:fig3}(a), at $\mathbf{q}=\mathbf{\Gamma}$ also have sizable
contribution due to their low energies.
Table~\ref{tab:table1} summarizes mode-resolved electron-phonon coupling
strength $\lambda_i$ for the three characteristic phonon modes
shown in Fig.~\ref{fig:fig3}(b).

Since $E_F\approx1{\text -}10\;\mathrm{meV}$ and
$\omega_{\mathrm{ph}}\approx10{\text -}200\;\mathrm{meV}$,
the adiabatic condition $\omega_\mathrm{ph}/E_F\ll1$ is extremely
violated in graphene moir\'e superlattices.
So the conventional McMillan formula does not apply.
Instead, superconducting transition temperature $T_c$ in the
nonadiabatic regime has nontrivial dependence on
the electronic bandwidth \cite{gorkov_2016,sadovskii_2019a}.
If we ignore the dispersion of phonon modes and the energy dependence of the electron DOS,
an explicit $T_c$ formula for the half-filled bands can be derived as \cite{sadovskii_2019a}
\begin{equation}
  T_c =
  \prod_i
  \left(\frac{\omega_i D}{\omega_i+D}\right)^{\lambda_i/\lambda}
  \!\!\!\!\exp\!\left(-\frac{1+\tilde{\lambda}}{\lambda-\mu^{*}}\right),
\end{equation}
where $D$ is the half bandwidth, $\omega_i$ and $\lambda_i$ are the energy and
electron-phonon coupling strength of the $i$th phonon mode,
$\tilde{\lambda}=2\sum_i \lambda_i D/(\omega_i+D)$ is the mass
renormalization constant, and
$\mu^{*}=\mu/(1+\mu\sum_i\ln(1+D/\omega_i)^{\lambda_i/\lambda})$
is the Coulomb pseudopotential.
We calculate $T_c$ at the half-filling Fermi energy
of the hole- (electron-) side flat bands for TBG and TTG (TDBG and TMBG).
Our results for $T_c$ are summarized in Table~\ref{tab:table1},
and show good agreement with experimental observations in TBG and TTG
\cite{cao_2018a,hao_2021a,park_2021}.

\begin{figure}
  \includegraphics[width=8.6cm]{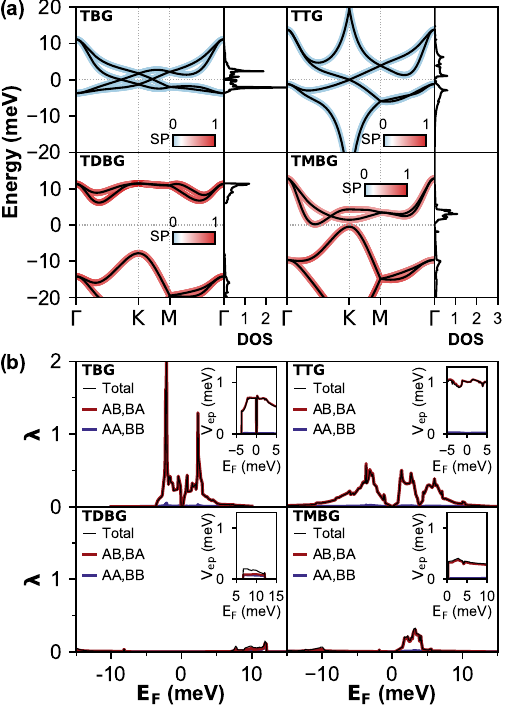}
  \caption {
    (a) Electronic structure and (b) electron-phonon coupling
    in (upper left) TBG with $\theta=1.08^\circ$,
    (upper right) TTG with $\theta=1.61^\circ$,
    (lower left) TDBG with $\theta=1.35^\circ$,
    and (lower right) TMBG with $\theta=1.25^\circ$
    under vertical electric field.
    The strength of electric field $E_z$ is 15 (5) mV/$\mathrm{\AA}$
    for TBG and TTG (TDBG and TMBG).
    The unit of DOS in (a) is states/spin/meV.
    In (a), colors of band lines represent the sublattice
    polarization. TBG and TTG have exactly zero sublattice polarization. In
    (b), black lines show the total coupling strength. Red (blue) lines represent
    contributions of inter- (intra-) sublattice electron-phonon couplings.
  }
  \label{fig:fig4}
\end{figure}

Lastly, we investigate electric-field effects on electronic structure
and electron-phonon coupling.
We consider a vertical electric field in our tight-binding Hamiltonian
by adding an electrostatic energy term $\Delta H = e E_z z$,
where $e>0$ is the elementary charge, $z$ is the $z$ coordinate of an atom,
and $E_z$ represents the total electric field consisting of external and
induced electric fields.

Figures~\ref{fig:fig4}(a) and \ref{fig:fig4}(b) show
electronic structures and electron-phonon coupling strengths
under the vertical electric field, respectively.
In our calculation,
TBG is nearly insensitive to the vertical electric field, except that two Dirac
points at $K$ are split in energy because of the potential energy difference between
two layers. Consequently, electron-phonon coupling strength is hardly affected
by the electric field.
In TTG, the electric field primarily affects highly dispersive monolayer Dirac
bands so that they are pushed away from the flat bands.
This makes the flat-band states more dispersive, reducing
the density of states and, accordingly, decreasing the electron-phonon coupling strength.
However, $V_{\mathrm{ep}}$ is not affected by electric fields.

On the other hand, the electronic structures in TDBG and TMBG are much more sensitive
to electric fields.
In both systems, flat bands are split into the electron and hole sides, and
the electron-side flat bands become narrower.
Nevertheless, TDBG under electric fields shows very weak $\lambda$ because
the sublattice polarization still suppresses $V_{\mathrm{ep}}$.
In more detail, we note a slight increase of $\lambda$ in the electron-side flat bands
of TMBG as the electric field polarizes the electron-side states
to the top monolayer where sublattice polarization is weaker.

In TBG and TTG, superconductivity often appears near spin-valley
flavor-polarized correlated phases at certain integer fillings \cite{wong_2020,zondiner_2020,hao_2021a,park_2021,park_2021b}.
Several studies have suggested that such correlated phases
compete with the superconductivity \cite{cao:2021,lian:2021}.
In particular, experiments with controlled metallic gates have shown that
superconductivity survives even after correlated phases disappear as a
result of the enhanced screening from metallic gates \cite{young:2020}.
Flavor polarization, if any, may weaken phonon-mediated superconductivity
in two ways.
First, flavor polarization may raise or lower band energies depending on
their flavors, increasing the band width and reducing the density
of states, which reduces the overall electron-phonon coupling strength.
Second, flavor polarization may lift energy degeneracy of
electron states with opposite momenta and opposite spins, for instance,
$|\bf{k},\uparrow\rangle$ and $|-\bf{k},\downarrow\rangle$,
which disturbs Cooper-pair formation for spin-singlet superconductivity.
Effects of flavor polarization and electron correlation need to be studied
for a full phase diagram of twisted graphene layers.
In addition, further studies should include doping-dependent band dispersions
and more accurate description of Coulomb matrix elements
\cite{guinea_2018,lewandowski_2021,cea_2021,scheurer_2020a,scheurer_2020b}.

To summarize,
we have studied electronic structure and electron-phonon coupling
in graphene moir\'e superlattices based on atomistic approaches.
We find that total electron-phonon coupling is strong for MA-TBG and MA-TTG,
but is an order of magnitude weaker for TDBG and TMBG,
where the Bernal-stacked layers induce sublattice polarization,
suppressing the nearest-neighbor electron-phonon matrix elements.
Our results provide a deeper understanding into the electron-phonon coupling
in graphene moir\'e superlattices, showing a correlation between strong
electron-phonon coupling and experimental observations of robust superconductivity.

\begin{acknowledgments}
  This work was supported by NRF of Korea (Grants No. 2020R1A2C3013673 and
  No. 2017R1A5A1014862) and KISTI
  supercomputing center (Project No. KSC-2020-CRE-0335). Y.W.C. acknowledges
  support from NRF of Korea (Global Ph.D. Fellowship Program
  NRF-2017H1A2A1042152).
\end{acknowledgments}

%\bibliographystyle{aps.bst}
%\bibliography{references}

%%%%% SUPPLEMENTAL MATERIAL
\onecolumngrid
\pagebreak
\setcounter{equation}{0}
\setcounter{figure}{0}
\setcounter{table}{0}
\renewcommand{\theequation}{S\arabic{equation}}
\renewcommand{\thefigure}{S\arabic{figure}}
\renewcommand{\bibnumfmt}[1]{[S#1]}
\renewcommand{\citenumfont}[1]{S#1}

%\pagebreak

\begin{center}
  \textbf{\large Supplemental Material: \\
Dichotomy of Electron-Phonon Coupling in Graphene Moir\'e Flat Bands}\\[.2cm]
  Young Woo Choi and Hyoung Joon Choi$^*$\\[.1cm]
  {\itshape Department of Physics, Yonsei University, Seoul 03722, Korea\\}
\end{center}

\begin{center}
\setlength{\fboxrule}{0pt}

\fbox{\begin{minipage}{0.8\textwidth}
\hspace{5pt} In this supplemental material, we present
detailed descriptions of our atomistic calculation methods for
relaxed atomic structures, phonon spectra,
electronic structures,
and electron-phonon matrix elements,
which are also used in our previous works on
electron-phonon coupling in twisted bilayer graphene [S1]
and electronic structure in twisted double bilayer graphene [S2].
In addition, we also present
formulas for sublattice analysis of electron-phonon interaction.
\end{minipage}}
\end{center}

\twocolumngrid

\subsection{S1. Structural Relaxation of Twisted Graphene Layers}
The twist angle $\theta$ for a commensurate moir\'e supercell of graphene
layers is given by
\begin{equation}
  \cos\theta = \frac{M^2+N^2+4MN}{2(M^2+N^2+NM)}
\end{equation}
for a pair of integers $(M, N)$ defining the supercell.
In the case of twisted bilayer graphene (TBG), for instance,
we consider supercell lattice vectors
\begin{equation}
\begin{aligned}
\bm{t}^{(1)}_1 & =  N \bm{a}_1+ M \bm{a}_2, \\
\bm{t}^{(1)}_2 & =  -M \bm{a}_1+ (M+N) \bm{a}_2
\end{aligned}
\end{equation}
of the bottom layer and
\begin{equation}
\begin{aligned}
\bm{t}^{(2)}_1 & =  M \bm{a}_1+ N \bm{a}_2, \\
\bm{t}^{(2)}_2 & =  -N \bm{a}_1+ (M+N) \bm{a}_2
\end{aligned}
\end{equation}
of the top layer before twist.
Here $\bm{a}_1 = a(\sqrt{3}/2,-1/2)$ and
$\bm{a}_2 = a(\sqrt{3}/2,+1/2)$ are the primitive lattice vectors
of graphene, and $a=2.46$ \AA~is the lattice parameter.
Then, a commensurate supercell is formed by rotating the bottom and top layers
by $-\theta/2$ and $\theta/2$, respectively, so that
the lattice vectors of the supercell are
\begin{equation}
\begin{aligned}
\bm{t}_1&=R(-\theta/2)\bm{t}^{(1)}_1=R(\theta/2)\bm{t}^{(2)}_1, \\
\bm{t}_2&=R(-\theta/2)\bm{t}^{(1)}_2=R(\theta/2)\bm{t}^{(2)}_2.
\end{aligned}
\end{equation}
Here $R(\theta)$ is the clockwise rotation by $\theta$
around the hollow center of the graphene hexagon.

Other twisted graphene layers are generated in a similar way.
Alternating-twist trilayer graphene (TTG) is formed by adding a graphene
layer on the top of TBG, where the third layer is aligned with the first layer.
For twisted double bilayer graphene (TDBG),
two Bernal-stacked bilayers are twisted instead of
two monolayers.
Twisted mono-bilayer graphene (TMBG), then, is formed by removing the topmost layer
from TDBG, resulting in a monolayer twisted on the top of a Bernal-stacked
bilayer.

After generating rigidly twisted graphene layers, we relax
atomic positions by minimizing the total energy as a function of
all atomic positions.
The total energy $U$ per a moir\'e supercell consists of the in-plane elastic
energy and interlayer van der Waals binding energy:
\begin{eqnarray}
    U & = & \frac{1}{2} \sum_{l} \sum_{i\alpha,pj\beta}
    C^\text{MLG}_{0i\alpha,pj\beta} \;
    \Delta\tau^{l}_{0i\alpha} \Delta\tau^{l}_{pj\beta} \nonumber \\
    & + & \frac{1}{2} \sum_{l\neq l'} \sum_{i,pj}
    V_\text{KC} ( \bm{\tau}^{l}_{0i}-\bm{\tau}^{l'}_{pj})~.
  \label{eq:etot}
\end{eqnarray}
Here $\bm{\tau}^{l}_{pi}=\bm{R}_p + \bm{\tau}^{l}_{i}$
is the position of the $i$th atom in the $l$th layer in
the $p$th moir\'e supercell located at $\bm{R}_p$,
$\alpha$ is the cartesian index ($\alpha=x,y,z$),
$\Delta \bm{\tau^{l}}_{pi} =
\bm{\tau}^{l}_{pi}-\bm{\tilde{\tau}}^{l}_{pi}$
is the deviation from the nonrelaxed position $\bm{\tilde{\tau}}^{l}_{pi}$,
and $C^\text{MLG}_{pi\alpha,p'j\beta}$ are force constants
between two atoms in the same layer up to the fourth-nearest neighbors.
The explicit form of the force contants is given by Eq.~(5) of
Ref.~S3 and their values
are shown in the column `4NNFC diagonal fit to GGA' of Table~3
of Ref.~S3,
obtained by fitting to {\em ab initio} phonon dispersion
calculations of monolayer graphene (MLG).
The interlayer binding energy is
calculated using Kolmogorov-Crespi (KC) potential
$V_\text{KC}$ that depends on interlayer atomic registry~[S4].
The explicit form of the KC potential is given by Eq.~(3) of
Ref.~S4, where values of parameters are
$C_0=15.71$~meV, $C_2=12.29$~meV, $C_4=4.933$~meV, $C=3.030$~meV,
$\delta =0.578$~{\AA}, $\lambda=3.629$~{\AA}$^{-1}$, $A=10.238$~meV, and
$z_0=3.34$~{\AA}, as given in Ref.~S4.
We use a conjugate gradient method to minimize the total energy $U$ as a function of all atomic coordinates within
the moir\'e supercell, whose total degrees of freedom are about 30 000
in the case of TBG with $\theta\sim1^\circ$.

\subsection{S2. Atomistic Calculation of Phonons in Twisted Graphene Layers}
We calculate atomic force constants
by taking the second derivatives of Eq.~(\ref{eq:etot})
with respect to atomic positions:
\begin{equation}
C^{l,l'}_{pi\alpha,p'j\beta} =
\partial^2 U / \partial \tau^{l}_{pi\alpha} \partial \tau^{l'}_{p'j\beta}.
\end{equation}
As the in-plane elastic energy is quadratic, the in-plane force constants are
just those of monolayer graphene,
\begin{equation}
C^{l=l'}_{pi\alpha,p'j\beta}=C^\text{MLG}_{pi\alpha,p'j\beta},
\end{equation}
which are not changed by lattice relaxation.
The interlayer force constants are the second derivatives of the KC potential,
\begin{equation}
C^{l\neq l'}_{pi\alpha,p'j\beta}=-\frac{\partial V_{\text{KC}}}{\partial x_\alpha \partial x_\beta}( \bm{\tau}^{l}_{pi}-\bm{\tau}^{l'}_{pj}),
\end{equation}
which are evaluated at relaxed atomic positions.

From the force constants, the dynamical matrix is
\begin{equation}
D_{i\alpha,j\beta}(\bm{q})= \sum_{p}
e^{i\bm{q}\cdot\bm{R}_p} \; C_{0i\alpha, pj\beta}/M_C
\end{equation}
for a phonon wave vector $\bm{q}$, where $M_C$ is the mass of a carbon atom.
Then, we solve the phonon eigenvalue problem
\begin{equation}
\omega^2_{\bm{q}\nu} \; e_{\bm{q}\nu,i\alpha} =
\sum_{j,\beta} D_{i\alpha,j\beta}(\bm{q}) \;
e_{\bm{q}\nu,j\beta}
\end{equation}
for the irreducible Brillouin zone to get the phonon energy $\omega_{\bm{q}\nu}$
and polarization vector $\bm{e}_{\bm{q}\nu,i}$ of the $\nu$th phonon mode.
The phonons in the rest of the Brillouin zone are obtained from the symmetry relations~[S5].
We use the ELPA library to diagonalize the dynamical matrix efficiently~[S6].

\subsection{S3. Atomistic Tight-Binding Method for Electronic Structure}
We use a single-orbital tight-binding approach to calculate electronic structures of twisted graphene layers.
Hamiltonian matrix elements are
\begin{equation}
  \label{h_tb}
    \hat{H} = \sum_{pi,p'j}
    t(\bm{\tau}_{pi}-\bm{\tau}_{p'j})
    |\phi_i;\bm{R}_p\rangle \langle\phi_{j};\bm{R}_{p'}|~,
\end{equation}
where $|\phi_i; \bm{R}_p\rangle$ is a carbon $p_z$-like orbital centered at
$\bm{\tau}_{pi}$. Here we drop the layer index on $\bm{\tau}_{pi}$
so the index $i$ sweeps all atoms in all layers in the $p$th moir\'e supercell
at $\bm{R}_p$.
We use the Slater-Koster-type hopping integral
\begin{eqnarray}
t(\bm{d}) &=& V_{pp\pi}^0 e^{-(d-a_0)/\delta} \{ 1-(d_z/d)^2 \} \nonumber \\
    & + &  V_{pp\sigma}^0 e^{-(d-d_0)/\delta}(d_z/d)^2,
\end{eqnarray}
where $\bm{d}$ is the displacement vector between two orbitals.
The hopping energy $V^0_{pp\pi}=-2.7\,\text{eV}$ is between in-plane nearest
neighbors separated by $a_0=a/\sqrt{3}=1.42\,\text{\AA}$, and
$V^0_{pp\sigma}=0.48\,\text{eV}$ is between two
vertically aligned atoms at the distance $d_0 = 3.35\,\text{\AA}$.
Here $\delta = 0.184 a$ is chosen to set the magnitude of
the next-nearest-neighbor hopping amplitude to be $0.1V^0_{pp\pi}$~[S7,S8].
We use the cutoff distance
$d_c = 10\,\text{\AA}$, beyond which the hopping integral is negligible.

\subsection{S4. Electron-Phonon Matrix Elements}
The total electron-phonon coupling strength $\lambda$ is
\begin{subequations}
\begin{eqnarray}
\label{eq:4a}
  \lambda_{n\bm{k}} &=& 2 N_F \sum_{m\bm{q}\nu}
  \frac{|g_{mn\nu}(\bm{k},\bm{q})|^2}{\omega_{\bm{q}\nu}} W_{m\bm{k+q}}, \\
  \lambda &=& \sum_{n\bm{k}} \lambda_{n\bm{k}} W_{n\bm{k}},
\label{eq:4b}
\end{eqnarray}
\end{subequations}
where $N_F$ is the electron density of states per spin at the Fermi energy
$E_F$, $W_{n\bm{k}}=\delta(E_F - \varepsilon_{n\bm{k}})/N_{F}$ is the
partial weight of the density of states contributed by
the $n$th electronic state $|n\bm{k}\rangle$
with wave vector $\bm{k}$ and
energy $\varepsilon_{n\bm{k}}$, and
$g_{mn\nu}(\bm{k},\bm{q})=\langle m\bm{k+q} |\delta_{\bm{q}\nu} \hat{H}
|n\bm{k}\rangle$ is the electron-phonon matrix element
between $|n\bm{k}\rangle$ and $|m\bm{k+q}\rangle$
mediated by the $\nu$th phonon mode of wave vector $\bm{q}$ and frequency $\omega_{\bm{q}\nu}$.
We calculate $W_{n\bm{k}}$ using the linear tetrahedron method~[S9].
In our tight-binding approach, we expand electron and phonon eigenstates in localized basis sets
to obtain the electron-phonon matrix element~[S1]
\begin{eqnarray}
  g_{mn\nu}(\bm{k},\bm{q})
   &= & l_{\bm{q}\nu} \sum_{i \alpha} e_{\bm{q}\nu,i \alpha}
     \sum_{pj} \frac{\partial}{\partial x_\alpha}
       t(\bm{\tau}_{0i} - \bm{\tau}_{pj}) \nonumber \\
    & \times & \{ e^{i\bm{k}\cdot\bm{R}_{p}}  c^*_{m\bm{k+q},i}  c_{n\bm{k},j}
\nonumber \\
 &&  + e^{-i(\bm{k+q})\cdot\bm{R}_{p}}  c^*_{m\bm{k+q},j} c_{n\bm{k},i} \},
\label{eq:gmnu}
\end{eqnarray}
where $l_{\bm{q}\nu}=\sqrt{\hbar/(2M_C\omega_{\bm{q}\nu})}$ is the length scale
of phonon mode $(\bm{q}\nu)$ and $c_{n\bm{k},i}$ is
the tight-binding coefficient of the electron state $|n\bm{k}\rangle$.
The Eliashberg function can also be calculated from the matrix elements as
\begin{eqnarray}
\alpha^2F(\omega) & = &  N_F \!\! \sum_{nm\nu\bm{k}\bm{q}}
|g_{mn\nu}(\bm{k},\bm{q})|^2  \nonumber \\
& \times &
W_{\!n\bm{k}} W_{\!m\bm{k+q}} \delta(\omega-\omega_{\bm{q}\nu}),
\end{eqnarray}
which shows the energy dependence of the momentum-averaged electron-phonon coupling.

\subsection{S5. Formulas for Sublattice Analysis of Electron-Phonon Coupling}

In Eq.~(\ref{eq:gmnu}), each term in the summation depends on two
atomic sites indexed by $i$ and $j$, so
the matrix element can be decomposed into intra- and inter-sublattice
contributions depending on whether $i$ and $j$ sites belong to the same
or different sublattices,
\begin{eqnarray}
  g_{mn\nu}(\bm{k},\bm{q}) & = &
  \underbrace{g^{AA}_{mn\nu}(\bm{k},\bm{q})+g^{BB}_{mn\nu}(\bm{k},\bm{q})}_{\text{intra-sublattice}}  \nonumber \\
 & + & \underbrace{g^{AB}_{mn\nu}(\bm{k},\bm{q})+g^{BA}_{mn\nu}(\bm{k},\bm{q})}_{\text{inter-sublattice}},
\end{eqnarray}
where
\begin{eqnarray}
 g^{S_1S_2}_{mn\nu}(\bm{k},\bm{q})
 & = &  l_{\bm{q}\nu} \!\!\! \sum_{i \in S_1, j\in S_2 }
      \sum_{\alpha,p}
       e_{\bm{q}\nu,i \alpha}
       \frac{\partial}{\partial x_\alpha}
       t(\bm{\tau}_{0i} - \bm{\tau}_{pj}) \nonumber \\
       & \times & \{
  e^{i\bm{k}\cdot\bm{R}_{p}}  c^*_{m\bm{k+q},i}  c_{n\bm{k},j} \nonumber \\
 &&  + e^{-i(\bm{k+q})\cdot\bm{R}_{p}}  c^*_{m\bm{k+q},j} c_{n\bm{k},i} \}
\end{eqnarray}
is the matrix element between sublattices $S_1$ and $S_2$.

In terms of the decomposed matrix elements,
we can also identify sublattice-dependent contributions to the total coupling strength,
\begin{equation}
\lambda = \!
\underbrace{\lambda^{AA}\!+\lambda^{BB}}_{\text{intra-sublattice}}
\!+\underbrace{\lambda^{AB}\!+\lambda^{BA}}_{\text{inter-sublattice}}
\!+\lambda^{\text{mixed}},
\end{equation}
where
\begin{equation}
\lambda^{S_1S_2} = 2N_F\sum_{nm\bm{k}\bm{q}\nu} \frac{\left|g^{S_1S_2}_{mn\nu}(\bm{k},\bm{q})\right|^2}
{\omega_{\bm{q}\nu}} W_{n\bm{k}} W_{m\bm{k}+\bm{q}}
\end{equation}
is the coupling strength between sublattices $S_1$ and $S_2$,
and the mixed term contains the rest of the matrix elements
$\lbrace g^{S_1S_2}_{mn\nu}(\bm{k},\bm{q}) \rbrace^* g^{S'_1S'_2}_{mn\nu}(\bm{k},\bm{q})$
with $(S_1,S_2)\neq(S'_1,S'_2)$.
As shown in Fig.~2(b) of the main text, the inter-sublattice terms $\lambda^{AB/BA}$
have the dominant contributions to the total coupling strength.
The intra-sublattice terms $\lambda^{AA/BB}$ are an order of magnitude weaker than $\lambda^{AB/BA}$.
We note that $\lambda^{\text{mixed}}$ is not always positive-valued, but its
magnitude is only comparable to $\lambda^{AA/BB}$ so it also has negligible contribution to the total coupling strength.

\vspace{.7cm}

\noindent
$^*$ h.j.choi@yonsei.ac.kr

\vspace{.2cm}

\hangindent=.7cm\hangafter=1\noindent
[S1]\hspace{.1cm}Y.~W. Choi and H.~J. Choi,
Strong electron-phonon coupling, electron-hole asymmetry, and nonadiabaticity
in magic-angle twisted bilayer graphene,
Phys. Rev. B \textbf{98}, 241412(R) (2018).

\hangindent=.7cm\hangafter=1\noindent
[S2]\hspace{.1cm}Y.~W. Choi and H.~J. Choi,
Intrinsic band gap and electrically tunable flat bands in twisted double
bilayer graphene,
Phys. Rev. B \textbf{100}, 201402(R) (2019).

\hangindent=.7cm\hangafter=1\noindent
[S3]\hspace{.1cm}L.~Wirtz and A.~Rubio,
The phonon dispersion of graphite revisited,
Solid State Commun.~\textbf{131},~141~(2004).

\hangindent=.7cm\hangafter=1\noindent
[S4]\hspace{.1cm}A.~N. Kolmogorov and V.~H. Crespi,
Registry-dependent interlayer potential for graphitic systems,
Phys. Rev. B \textbf{71}, 235415 (2005).

\hangindent=.7cm\hangafter=1\noindent
[S5]\hspace{.1cm}A. A. Maradudin and S. H. Vosko,
Symmetry properties of the normal vibrations of a crystal,
Rev. Mod. Phys. {\bf 40}, 1 (1968).

\hangindent=.7cm\hangafter=1\noindent
[S6]\hspace{.1cm}A. Marek, V. Blum, R. Johanni, V. Havu, B. Lang, T. Auckenthaler, A. Heinecke, H.-J. Bungartz, and H. Lederer,
The ELPA Library - Scalable Parallel Eigenvalue Solutions for Electronic Structure Theory and Computational Science,
J. Condens. Matter Phys. {\bf 26}, 213201 (2014).

\hangindent=.7cm\hangafter=1\noindent
[S7]\hspace{.1cm}T.~Nakanishi and T.~Ando,
Conductance of Crossed Carbon Nanotubes,
J. Phys. Soc. Japan \textbf{70}, 1647 (2001).

\hangindent=.7cm\hangafter=1\noindent
[S8]\hspace{.1cm}P.~Moon and M.~Koshino,
Energy spectrum and quantum Hall effect in twisted bilayer graphene,
Phys. Rev. B \textbf{85}, 195458 (2012).

\hangindent=.7cm\hangafter=1\noindent
[S9]\hspace{.1cm}P. E. Bl\"{o}chl, O. Jepsen, and O. K. Andersen,
Improved tetrahedron method for Brillouin-zone integrations,
Phys. Rev. B {\bf 49}, 16223 (1994).

\end{document}